\documentclass{article}
\usepackage{spconf,amsmath,graphicx,hyperref}

\usepackage{amsmath,amssymb,amsfonts}
\usepackage{algorithmic}
\usepackage{textcomp}
\usepackage{hhline}
\usepackage{multirow,array}
\usepackage{xcolor}

\newcommand{\olsi}[1]{\,\overline{\!{#1}}} 
\usepackage{comment}

\title{Acoustic Non-Stationarity Objective Assessment with Hard Label Criteria for Supervised Learning Models}
\name{Guilherme Zucatelli, Ricardo Barioni, Gabriela Dantas \thanks{This work was funded by Samsung Eletrônica da Amazonia Ltda., under the auspices of the Brazilian Federal Law of Informatics no. 8248/91.}}
\address{Speech Processing Team \, - \, SiDi - Intelligence \& Innovation Center \, - \, São Paulo, Brazil
        }

\begin{document}
\maketitle
\begin{abstract}
Objective non-stationarity measures are resource intensive and impose critical limitations for real-time processing solutions.
In this paper, a novel Hard Label Criteria (HLC) algorithm is proposed to generate global non-stationarity labels for acoustic signals, enabling supervised learning strategies to be trained as stationarity estimators.
{\color{black}The HLC is first evaluated on state-of-the-art general-purpose acoustic models, demonstrating that these models capture stationarity information.}
Furthermore, the first-of-its-kind HLC-based Network for Acoustic Non-Stationarity Assessment (NANSA) is proposed.
{\color{black} NANSA models outperform competing approaches, achieving up to 99\% classification accuracy, while solving the computational infeasibility of traditional objective measures.}
\end{abstract}

\section{Introduction}
Acoustic signals are commonly considered non-stationary across various research domains, including automatic speech recognition (ASR) \cite{shin2024statistical}, computational auditory scene analysis (CASA) \cite{wang2006computational}, and speech enhancement (SE) \cite{ristea2025icassp, cohen_2001}.
However, despite the usual assumption, experiments are rarely accompanied by objective assessments, which are essential to validate the hypothesis and evaluate strategies under different degrees of temporal and spectral variations.

One objective non-stationarity measure successfully applied in the acoustic domain is the Index of Non-Stationarity (INS) \cite{flandrin_2018, flandrin_10}. 
The INS has been used in contexts related to audio synthesis and adaptive learning \cite{zucatelli_2021}, speech intelligibility improvement \cite{zucatelli_2020}, emotion recognition \cite{vieira_2020} and  acoustic source classification \cite{zucatelli_2019}.
Nevertheless, INS faces major computational limitations for real-time applications due to resource-intensive steps, such as generating stationary synthetic references and performing multi-scale spectral comparisons. 
Finally, INS lacks an objective criterion for labeling an entire signal, often requiring expert interpretation of statistical outputs—a process that is labor-intensive and impractical at scale or on resource-constrained devices.

In this paper, we address the computational drawbacks of INS by proposing a novel Hard Label Criteria (HLC) algorithm to provide a global and objective assessment of non-stationarity in acoustic signals.
Unlike traditional INS, the proposed HLC evaluates stationarity over \emph{complementary regions}, producing a single binary label per signal.
This enables data-driven models to estimate non-stationarity as a binary classification task, transforming the previously demanding INS calculations into a simple inference process executable within milliseconds.

The HLC algorithm is first applied to fine-tune state-of-the-art general-purpose acoustic models PANNs \cite{pumbley_2020}, AST \cite{glass_2021}, and PaSST \cite{koutini_2022}.
As an additional contribution, we employ HLC to train a dedicated model: the Network for Acoustic Non-Stationarity Assessment (NANSA), along with its lightweight version, NANSA\textsubscript{LW}.
It is demonstrated that all acoustic models are reliable to HLC non-stationarity classification, with strong performances on AudioSet \cite{audioset_2017}, DCASE \cite{dcase_2018}, and FSD50K \cite{fsd50k_2021} datasets.
Notably, NANSA models surpass other approaches, achieving the best overall results.

\vspace{-.3cm}
\section{Proposed Method} 
\vspace{-.1cm}
\subsection{Review of the INS Framework}
\vspace{-.1cm}

The INS is a stationarity testing method relative to an \emph{observation scale}, applicable in both stochastic and deterministic contexts \cite{flandrin_10}. 
A key contribution of this work is the adoption of scale-relative INS to generate a global stationarity label, which serves as ground truth for training neural networks (see Section~\ref{sec:hard_label_criteria}).

The INS measures stationarity of a target signal $x(t)$ of length $T$ based on a spectral distance $\mathcal{D}$  and a family of $J$ surrogates $\{s_{j}(t), \ j=1, \dots, J\}$.
A surrogate is a theoretically stationary version of the original signal, forming the basis of the null hypothesis of stationarity \cite{flandrin_10}.
Each surrogate $s_{j}(t)$ is synthesized by modifying the spectral phase of $x(t)$ using the $j$-th realization of a uniform distribution $\mathcal{U}[-\pi, \pi]$.

\begin{figure}[t!]
    \centering
    \includegraphics[height=2.95cm]{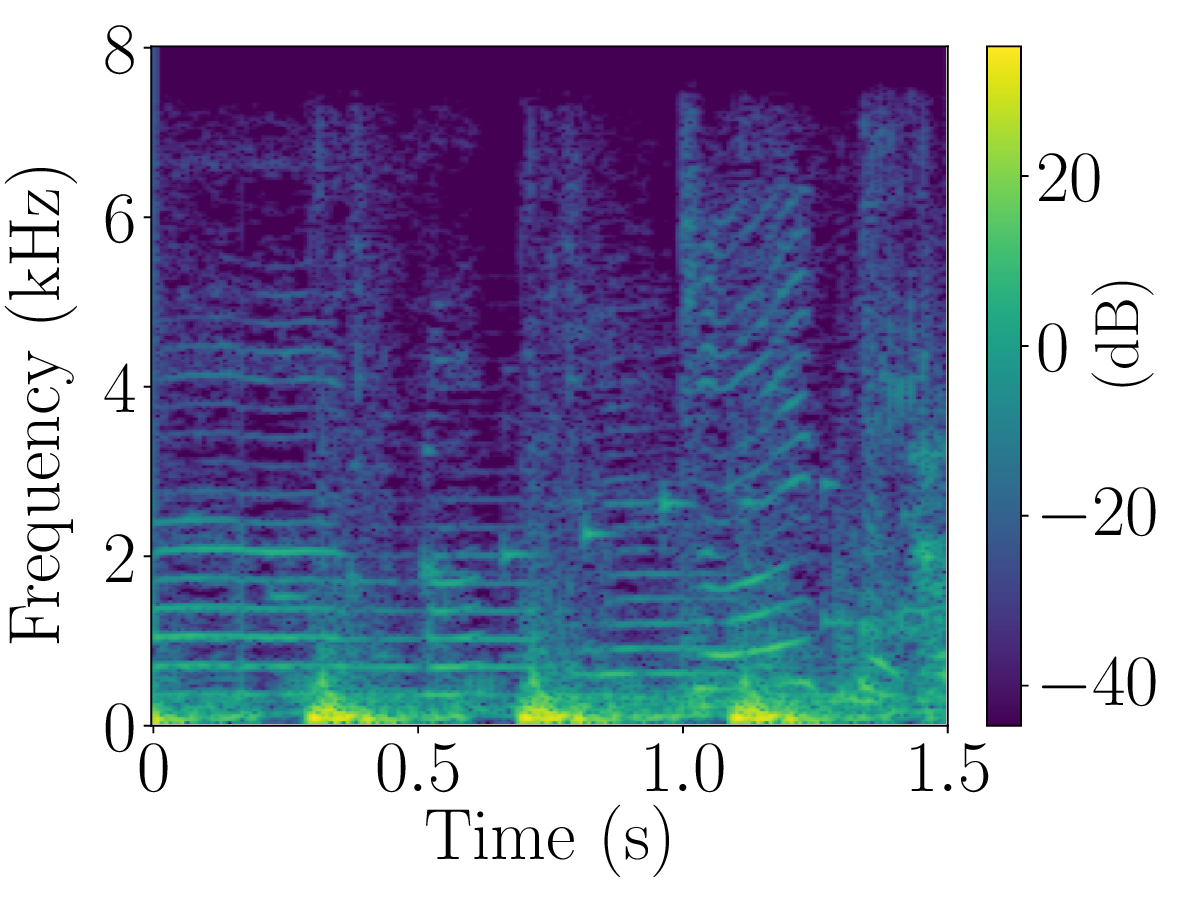}
    \includegraphics[height=2.95cm]{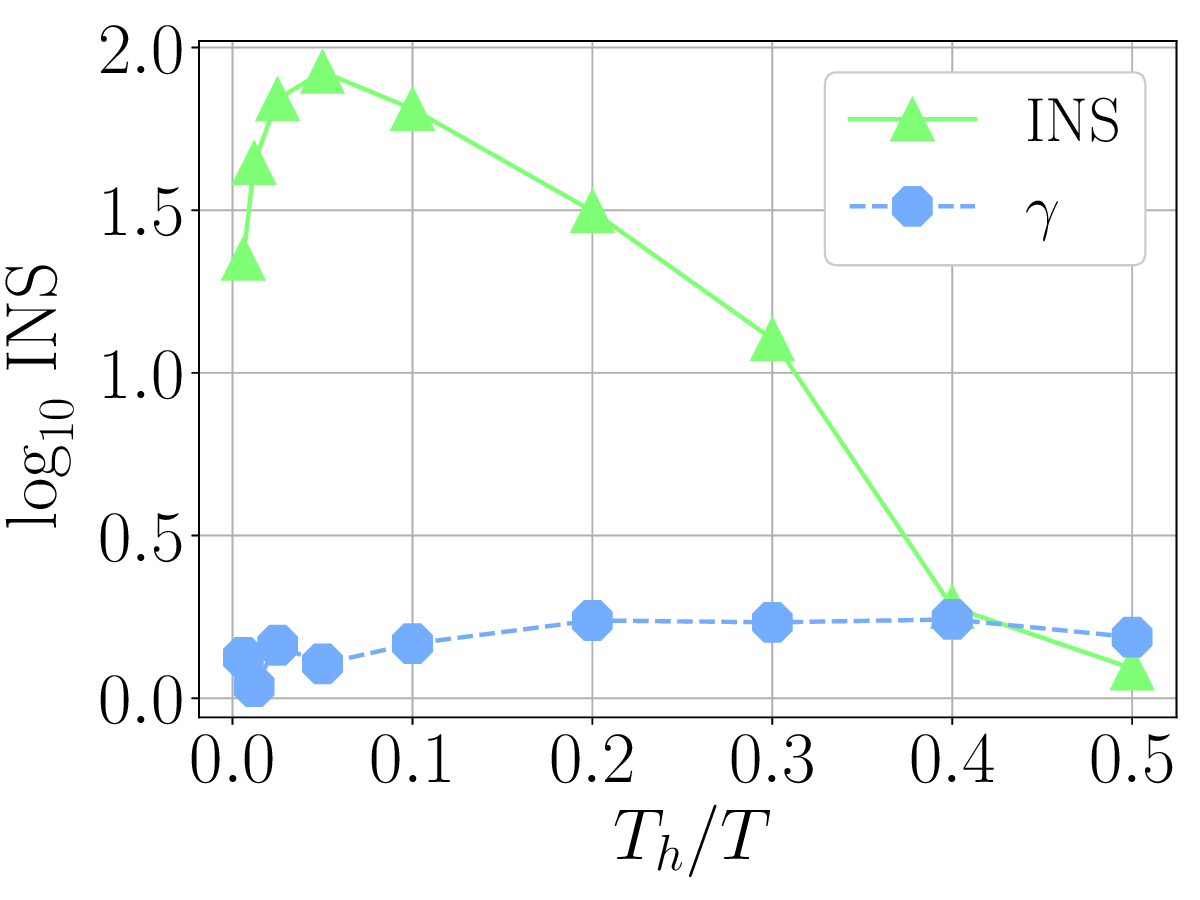}
    
    {\vspace{-0.5cm} \small (a)} \\
    
    \includegraphics[height=2.95cm]{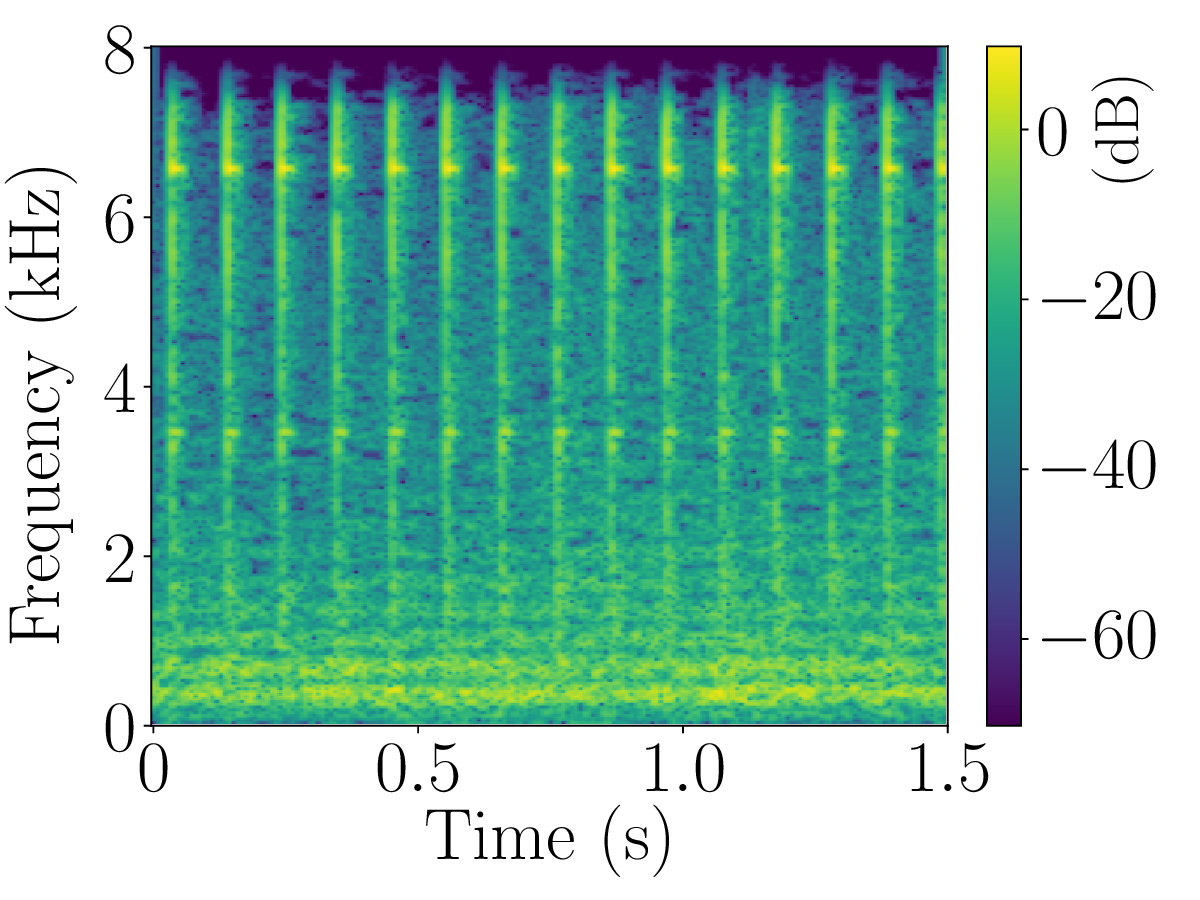}
    \includegraphics[height=2.95cm]{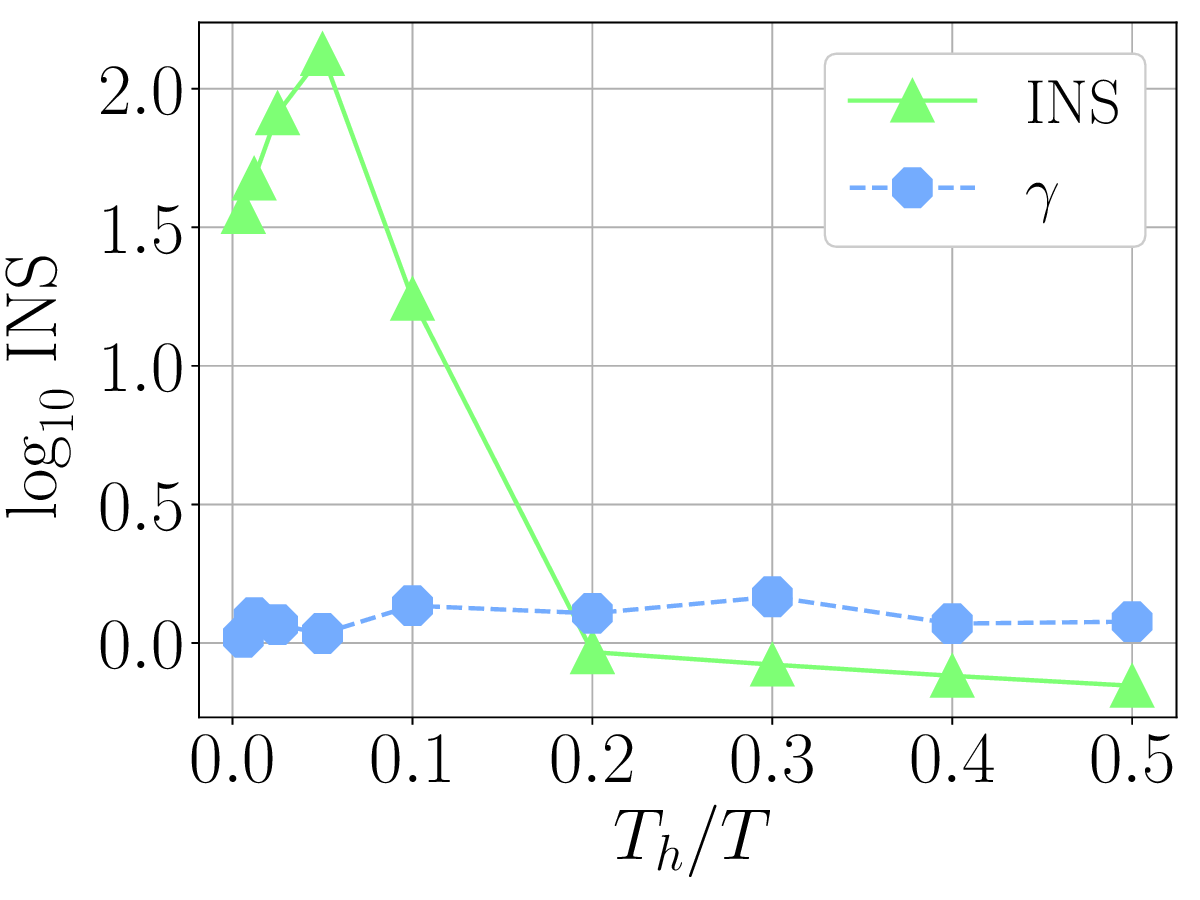}
    
    {\vspace{-0.5cm}\small (b)} \\
    
    \includegraphics[height=2.95cm]{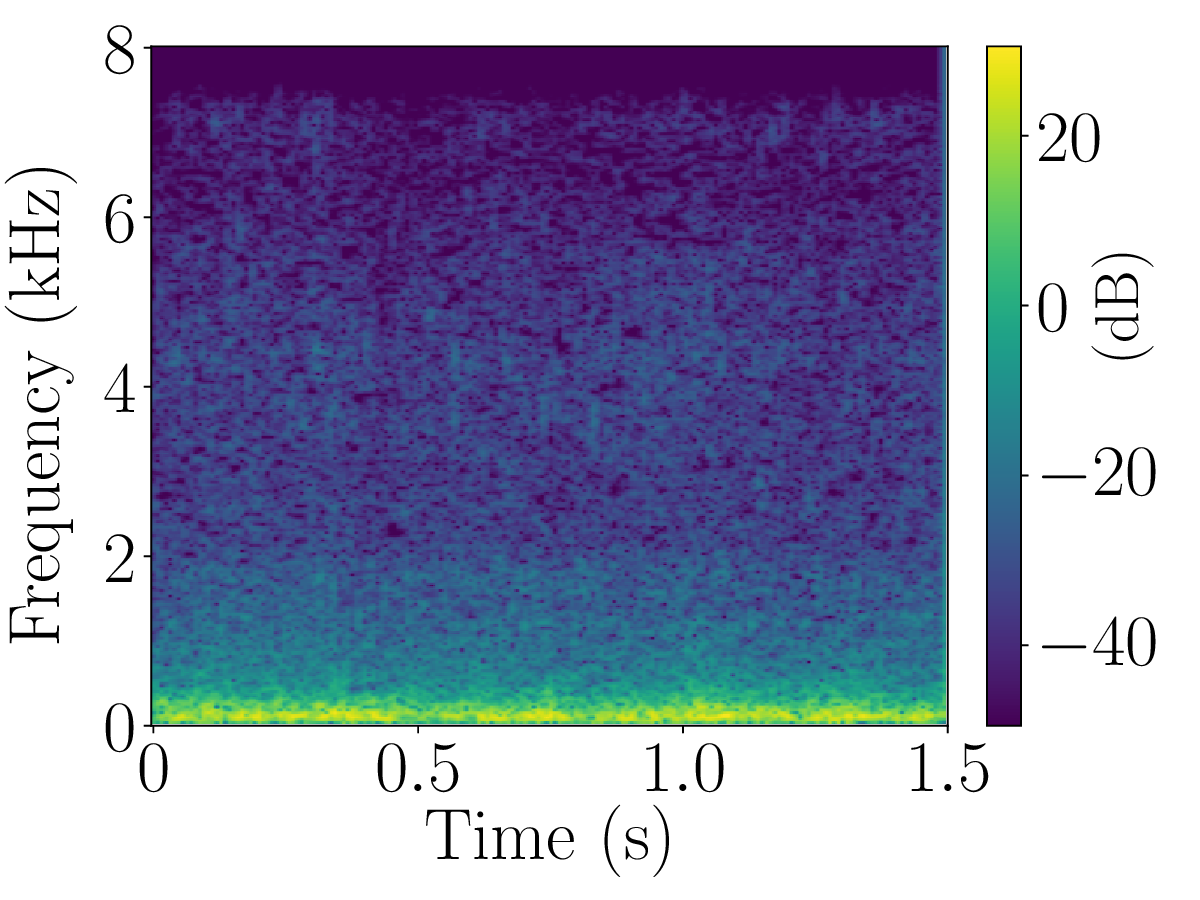}
    \includegraphics[height=2.95cm]{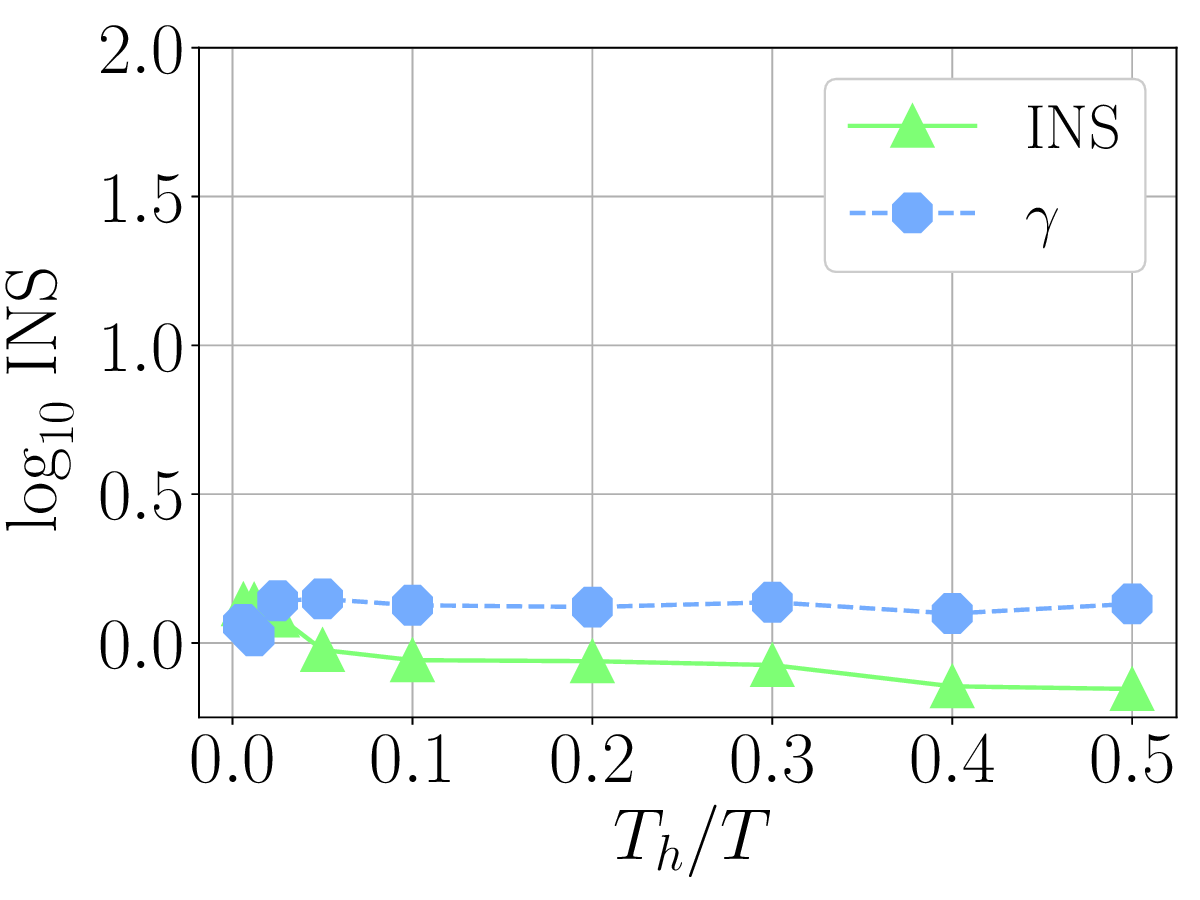}
    
    {\vspace{-0.5cm} \small (c)} \\
    
    \vspace{-0.2cm}
    \caption{Sample spectrogram signals and corresponding INS values extracted from AudioSet eval dataset: Noisy Speech (a), Wooden Knock (b) and Blowing Wind (c).}
    \label{fig:sig_spec_ins}
    \vspace{-0.5cm}
\end{figure}

Given the spectrograms of the target signal and of its surrogates $S_x(t_h, f)$ and $S_{s_j}(t_h, f)$ respectively, the dissimilarity between global and local frequency features is defined as
\vspace{-0.3cm}
\begin{equation}
    c_{z} := \mathcal{D}(\ S(t_h, \boldsymbol{\cdot}) \ , \langle \ S(t_h, \boldsymbol{\cdot}) \ \rangle_z, \ z=1, \dots, Z),
\end{equation}
where $\langle \ S(t_h, \boldsymbol{\cdot}) \ \rangle_z$ is the spectrogram of section $z$ for local observation window $T_h \le T/2$ and scale $T_h/T \in (0,0.5]$.

The dispersion of distances under the null hypothesis of stationarity can be characterized by the distribution of empirical variances $\{ \Theta_0(j) = \text{var}( c_z^{s_j} )_{z=1,\dots,Z} , j=1,\dots,J \}$, whereas the effective test is based on the statistics $\Theta_1 = \text{var}( c_z^{x} )_{z=1,\dots,Z}$.
The INS value is then computed as
\begin{equation}
    \label{eq:ins_equation}
     \text{INS} (T_h/T) := \sqrt{\frac{\Theta_1}{\langle \Theta_0(j) \rangle_j}},
\end{equation}
and a threshold $\gamma \approx 1$ is defined, such that the signal is non-stationary at scale $T_h/T$ when $\text{INS} (T_h/T)>\gamma$.

The INS implementation used in this work follows that of \cite{zucatelli_sbrt23}, where the spectral distance $\mathcal{D}$ is computed from a multi-taper spectral representation and defined as a combination of the log-spectral deviation and the Kullback–Leibler divergence, as described in \cite{flandrin_10}.

Fig.~\ref{fig:sig_spec_ins} depicts the spectrograms, INS values (in green), and stationarity thresholds $\gamma$ (in blue) for three 1.5-second samples from AudioSet \cite{audioset_2017}.
In the first example, the signal is classified as non-stationary for scales $T_h/T < 0.4$ and stationary otherwise.
That is, only segments with duration $T_h \geq 0.4T$ are sufficiently similar to the global spectrogram.
In the second case, a clear spectral pattern is observed, and the signal is non-stationary for $T_h/T < 0.2$, indicating that only shorter segments (less than 0.3 seconds) exhibit spectral distributions sufficiently distinct from the global pattern.
In the final example, the spectral energy distribution remains consistent over time, and the signal is stationary across all observable scales.

\begin{table}[t!]
\centering 
\renewcommand{\tabcolsep}{1.5mm}
\renewcommand{\arraystretch}{1.5}
\vspace{-0.2cm}
\caption{Correct HLC labelling for $1000$ random samples of acoustic sources from RSG-10 database.} 
\vspace{0.1cm}
\fontsize{9pt}{9pt}\selectfont
\begin{tabular}{|c c | c c c|} \hline 
    \multicolumn{2}{|c|}{ Stationary } & \multicolumn{3}{|c|}{ Non-Stationary} \\ \hline
    Office & Volvo & Babble & Factory & Machine Gun  \\ \hhline{-----} 
    95\% & 99\% & 100\% & 96\% & 99\%  \\ \hline
\end{tabular}
\label{tab:hlc_validation}
\vspace{-0.2cm}
\end{table}

\subsection{The Hard Label Criteria (HLC)}
\label{sec:hard_label_criteria}

In an intuitive analysis, the first and last signals of Fig.~\ref{fig:sig_spec_ins} could be globally categorized due to a common INS behavior for most scales.
However, that is not the case for the second example, which illustrates the necessity of a global objective assessment criterion for acoustic non-stationarity. 

The HLC algorithm is designed to estimate a single non-stationarity label per acoustic signal.
The proposed strategy relies on two steps: evaluating non-stationarity \emph{per region} and grouping these estimates into a universal label.

Let $\mathcal{T}$ be an ascending order sequence of observable scales $T_h/T$ divided into $K$ regions $\mathcal{T}_k$, such that $|\mathcal{T}_k|=N$, $\mathcal{T}_k \cap \mathcal{T}_{k'} = \varnothing$, $\ \forall \ k \ne k'$, and $\bigcup_{k=1}^{K} \mathcal{T}_k = \mathcal{T}$.
For notation simplicity, the elements of $\mathcal{T}_k$ will be denoted as $T_{kn}$, i.e., the $n$-th observable scale from the $k$-th region.
An adaptive threshold $\gamma_{HLC}$ for regions $\mathcal{T}_k$ is proposed as means to 
determine the subset $\mathcal{T}_k^{NS}$ of all scales $T_{kn} \in \mathcal{T}_k$ for which the signal is non-stationary, 
\begin{equation}
    \mathcal{T}^{NS}_k=\{T_{kn} \in \mathcal{T}_k \ : \ \text{INS}(T_{kn})>\gamma_{HLC}\}.
\end{equation}
Given the subset $\mathcal{T}_k^{NS}$, we introduce a binary function to characterize the non-stationarity of a region as
\begin{equation}
    f_\text{region}(\mathcal{T}_k) = \bigg\{
        \begin{matrix}
        1, & |\mathcal{T}^{NS}_k| > |\olsi{\mathcal{T}^{NS}_k}| \\
        0, & \text{otherwise}
        \end{matrix} \ \ . 
\end{equation}
The adaptive threshold is defined as $\gamma_{HLC} = \alpha_{HLC} \cdot \gamma$, where $\gamma$ is the INS stationarity threshold and $ \alpha_{HLC} > 1$ is an adjustable parameter.
Hence, $\gamma_{HLC} > \gamma$ imposes \emph{harder} (more restrictive) criteria over the stationary hypothesis, removing numerical outliers and establishing the stationarity condition over regions $\mathcal{T}_k$.

As a final step of HLC algorithm, the global label is obtained by the majority of non-stationary regions as
\begin{equation}
    \label{eq:hlc_func}
    f_\text{HLC}(\mathcal{T}_1, \dots,\mathcal{T}_K) = \bigg\{
        \begin{matrix}
        1, & \sum_{k=1}^{K} f_\text{region}(\mathcal{T}_k) > K/2  \\
        0, & \text{otherwise}
        \end{matrix} \ \ .
\end{equation}
Therefore, $f_\text{HLC}$ defines a single binary non-stationarity label based on all non-stationary regions (and observable scales) of a target acoustic signal.

The HLC algorithm is validated for acoustic signals extracted from RSG-10 database \cite{varga1993assessment}.
The sources are selected based on the physical interpretation of stationarity (Office and Volvo) and non-stationarity (Babble, Factory and Machine Gun), as in \cite{flandrin_10}.
Table \ref{tab:hlc_validation} shows the correct HLC  labeling for $1000$ random samples of each source.
The proposed algorithm attains an average accuracy of $98\%$, in accordance with the physical characterization of selected acoustic signals.

\subsection{NANSA Architecture and Training Criterion}
\begin{figure}[t!]
    \centering
    \includegraphics[width=\linewidth]{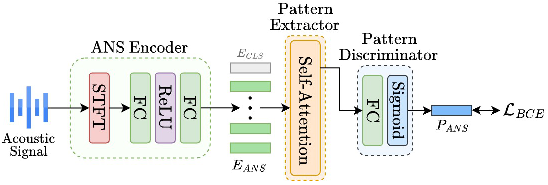}
    \vspace{-0.6cm}
    \caption{The NANSA model diagram.}
    \label{fig:model_diagram}
    \vspace{-0.5cm}
\end{figure}

As an additional contribution, the specialized Network for Acoustic Non-Stationary Assessment (NANSA) is proposed, which consists of three modules as illustrated in Fig.~\ref{fig:model_diagram}.

In the ANS Encoder, the Short-Time Fourier Transform (STFT) is applied every $20$~ms with 50\% overlap, at a 16~kHz sampling rate.
Resulting spectrogram $S \in \mathbb{R}^{T_\text{ANS} \times 257}$ is processed by two fully connected (FC) layers with scaling factors $\beta_{FC}$ and $1/\beta_{FC}$, separated by a ReLU activation, producing the embedding $E_{ANS}$.  
A classification embedding $E_{CLS}$ is appended to $E_{ANS}$.
The transformer-based Pattern Extractor uses self-attention to model both local and long-range temporal dependencies, enabling robust extraction of non-stationary patterns. Since the INS computation operates on spectrogram segments, unitary temporal patches and positional embeddings are employed \cite{waswani_2017}.
The probability $P_{ANS}$ is obtained from the first output embedding of this module.
Training is carried with binary cross-entropy loss $\mathcal{L}_{BCE}$, with ground truth labels provided by the $f_{HLC}$ function in (\ref{eq:hlc_func}).

The full NANSA model employs 11 self-attention layers, each with 3 heads and a 192-dimensional input.
Its lightweight variant, NANSA\textsubscript{LW}, uses 4 self-attention layers with 3 heads and a 64-dimensional input, targeting resource-constrained devices.

\vspace{-0.2cm}
\section{Experiments} 
\vspace{-0.1cm}
\subsection{Datasets and Baseline Models}

The acoustic non-stationarity classification is designed for supervised learning strategies based on HLC labels.
Experiments are conducted using signals from AudioSet \cite{audioset_2017}, DCASE \cite{dcase_2018} and FSD50K \cite{fsd50k_2021}.
These datasets are originally intended for acoustic sources, scenes and events classification, comprising a diverse collection of audio signals.
Standard dataset splits are used for training and evaluation.

Baseline state-of-the-art acoustic models are PANNs \cite{pumbley_2020}, AST \cite{glass_2021}, and PaSST \cite{koutini_2022}.
These publicly available general-purpose pretrained models are fine-tuned by replacing their final classification layers to perform the downstream non-stationarity classification task, while keeping all other parameters fixed.
Similar to the baselines, NANSA and NANSA\textsubscript{LW} are pretrained on the \emph{unbalanced} subset of AudioSet.

\begin{table*}[t!]
\centering 
\renewcommand{\tabcolsep}{1.0mm}
\renewcommand{\arraystretch}{1.2}
\caption{Comparison of competing supervised learning baseline models with proposed NANSA and NANSA\textsubscript{LW}: number of parameters, million MACs, acoustic non-stationarity classification Accuracy (\%), EER (\%) and F1 score. Lower values of EER, and higher values of Accuracy and F1 scores are better. Best results are presented in \textbf{bold}. } 
\begin{tabular}{l c c c c c c c c c c c c c} \hline 
    \multirow{2}{*}{ \shortstack[c]{Acoustic \\ Models} } & \multirow{2}{*}{ \# Params } & \multirow{2}{*}{ MMACs } & \multicolumn{3}{c}{AudioSet} & & \multicolumn{3}{c}{DCASE} & & \multicolumn{3}{c}{FSD50K}  \\ \hhline{~~~---~---~---} 
    & & & Acc (\%) & EER (\%) & F1 &  & Acc (\%) & EER (\%) & F1 & & Acc (\%) & EER (\%) & F1 \\ \hline
    PANNs \cite{pumbley_2020} & 81.04 M & 1736 & 90.82 & 9.25 & 0.925 &  & 98.27 & 6.37 & 0.578 & & 92.52 & 7.21 & 0.931 \\ \hhline{~~~~~~~~~~~} 
    AST \cite{glass_2021} & 94.04 M & 16785 & \bf 92.37 & \bf 7.92 & \bf 0.938 &  & 98.20 & 5.48 & 0.594 & & 93.86 & 6.26 & 0.943 \\  \hhline{~~~~~~~~~~~} 
    PaSST \cite{koutini_2022} & 83.35 M & 15021 & 92.02 & 8.24 & 0.936 &  & \bf 98.35 & \bf 5.26 &\bf  0.612 & &\bf  94.18 & \bf 5.80 &\bf  0.948 \\  \hline
    NANSA & 5.50 M & 585 & \bf 94.25 & \bf 5.87 & \bf 0.954 &  & \bf 99.01 & \bf 2.68 & \bf 0.801 & & \bf 95.41 & \bf 4.59 & \bf 0.958 \\  \hhline{~~~~~~~~~~~}
    {\textcolor[Gray]{10}{$-$ANS Encoder}} & 4.97 M & 505 & 93.52 & 6.58 & 0.948 &  & 98.84 & 2.91 & 0.748 & & 94.85 & 5.09 & 0.953 \\  \hhline{~~~~~~~~~~~}
    NANSA\textsubscript{LW} & 655.9 K & 88 & 93.27 & 6.73 & 0.946 &  & 98.89 & 2.91 & 0.780 & & 94.93 & 4.95 & 0.955 \\ \hhline{~~~~~~~~~~~}
    {\textcolor[Gray]{10}{$-$ANS Encoder}} & 126.3 K & 8 & 92.66 & 7.47 & 0.941 &  & 98.83 & 3.29 & 0.759 & & 94.39 & 5.62 & 0.949 \\ \hline
\end{tabular}
\label{tab:all_scores}
\vspace{-.3cm}
\end{table*}

\subsection{Implementation Details}
The INS assessment and label generation are computationally intensive and relied on the IARA Lab, one of the largest AI supercomputers in the world \cite{TOP500list}.
Further steps were carried on x86 Linux machines with NVIDIA V100 GPU.

Audio signals are segmented into 1.5-second clips, consistent with typical durations in speech and on-device audio applications \cite{shin_2022, warden_2018}.
Experiments are carried with HLC algorithm configured for $K=3$ regions $\mathcal{T}_k$ to capture short-, mid-, and long-term temporal dynamics.
The first two regions use approximately geometric progressions, with $\mathcal{T}_1=\{0.006, 0.012, 0.025\}$ ($9.0–37.5$ ms) and $\mathcal{T}_2=\{0.05, 0.1, 0.2\}$ ($75–300$ ms), while $\mathcal{T}_3=\{0.3, 0.4, 0.5\}$ ($400–750$ ms) is linear. This design is a reflection of higher observation scales exhibiting slower non-stationarity variations, as illustrated in Fig. \ref{fig:sig_spec_ins}. Additionally, $\alpha_{HLC}$ is conservatively set to $10$, i.e., HLC defines non-stationarity within a region by imposing a one-order-of-magnitude higher threshold across most INS observable scales.
All models are trained for 20 epochs, learning rate of $10^{-4}$ and Adam optimizer \cite{kingma2014adam}.

\vspace{-.2cm}
\subsection{Metrics and Statistical Analysis}
In line with other acoustic classification tasks \cite{pumbley_2020, glass_2021, koutini_2022}, model performance is primarily evaluated using accuracy.
Additionally, Equal Error Rate (EER) and F1-score are reported, as they are standard for imbalanced binary classification.
Receiver Operating Characteristic (ROC) curves and Area Under the Curve (AUC) scores are also provided.
To validate the significance of results, we employ the pairwise statistical testing method from \cite{bengio_2004}.

\section{Results and Discussion}
Table~\ref{tab:all_scores} presents the classification accuracy, EER, and F1-score for HLC-based acoustic non-stationarity assessment.
All baseline models (PANNs, AST, and PaSST) achieve over 90\% accuracy, indicating that general-purpose acoustic models are capable of capturing non-stationarity information.
Among them, the attention-based AST and PaSST outperform PANNs in both accuracy and F1-score, while achieving lower EER values.
However, these improvements come with significantly higher memory and compute (MMAC) costs, which are up to an order of magnitude greater than PANNs.

\begin{figure}[t!]
    \centering
    {\bf \small \hspace{0.3cm} AudioSet \hspace{2.9cm} DCASE} \\
    \hspace{-.35cm}
    \includegraphics[width=4.1cm]{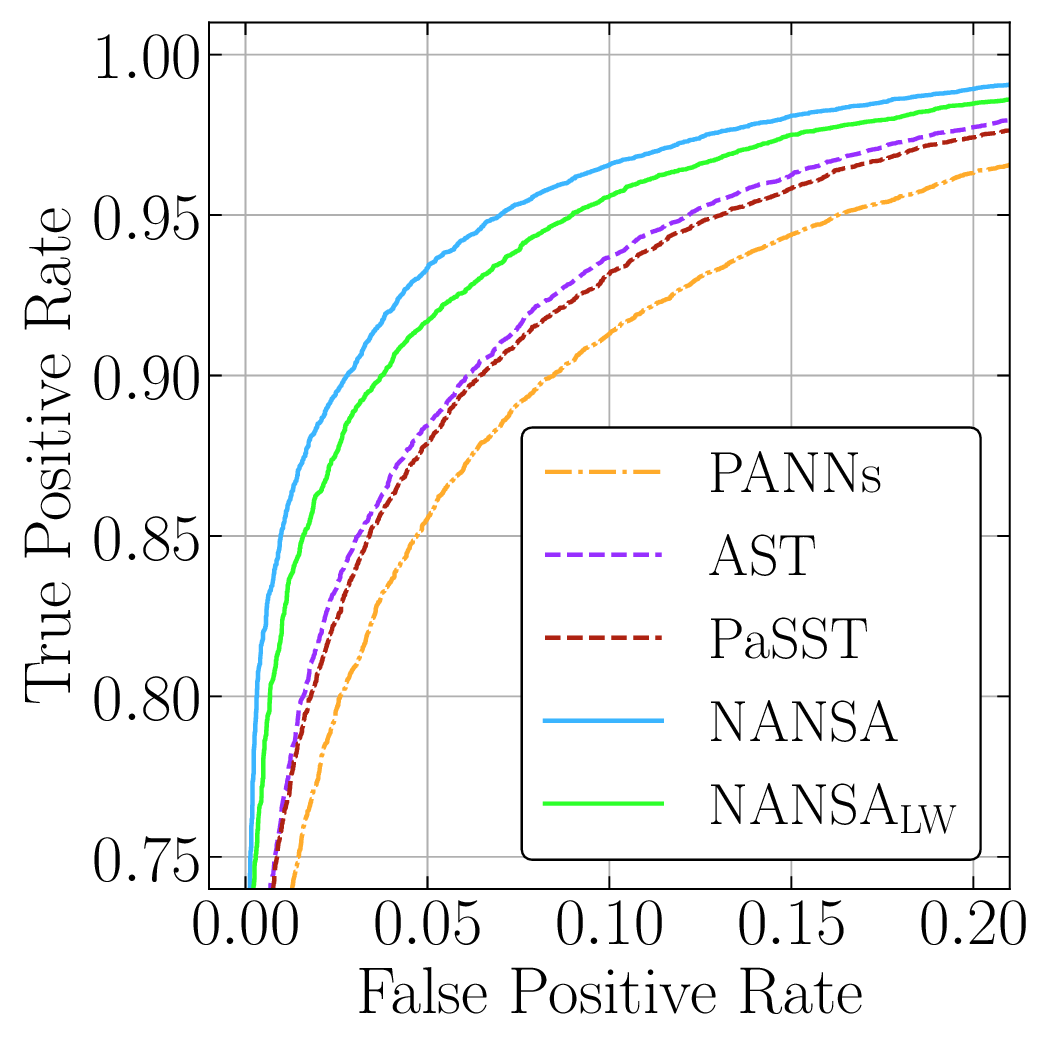}
    \includegraphics[width=4.1cm]{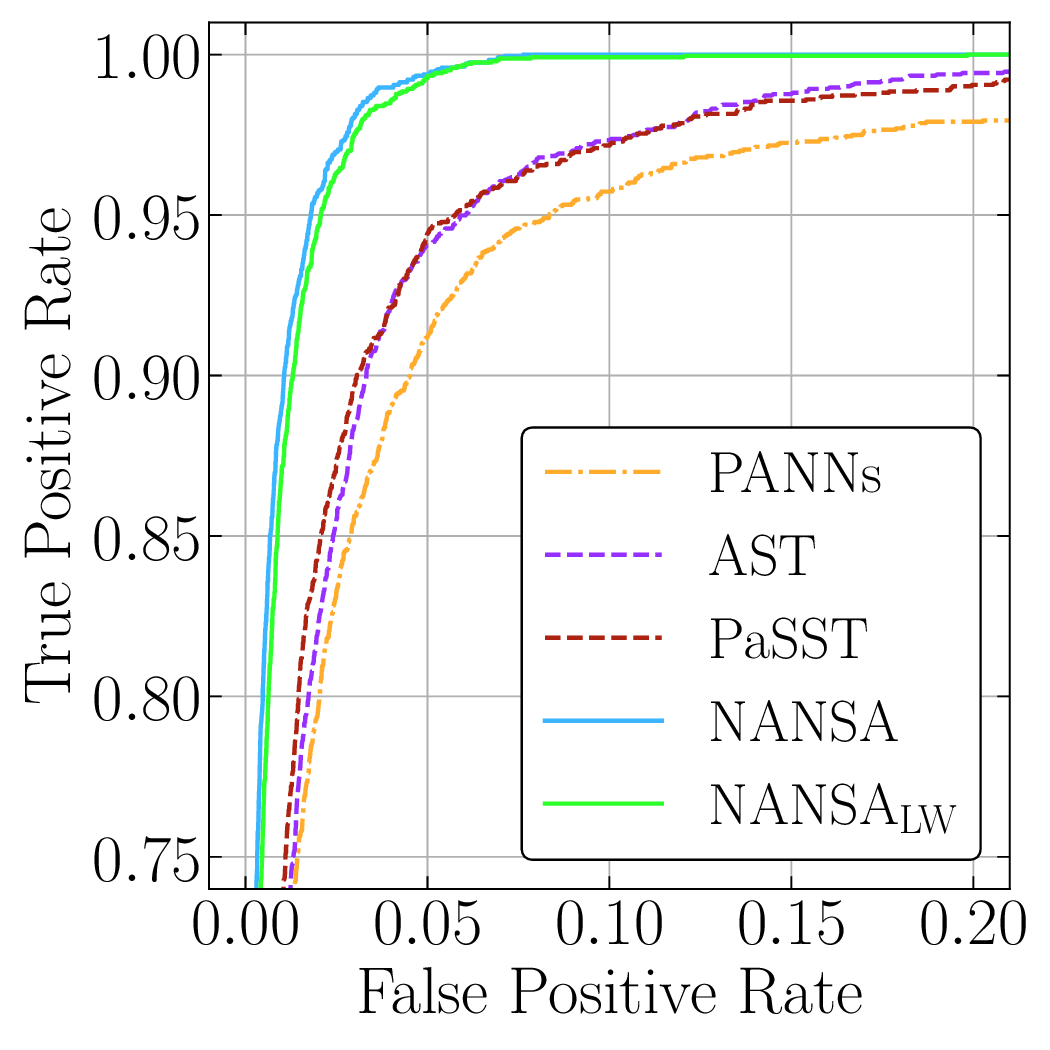}
    {\bf \small \hspace{0.0cm} FSD50K \hspace{3.6cm} } \\
    \hspace{-.35cm}
    \includegraphics[width=4.1cm]{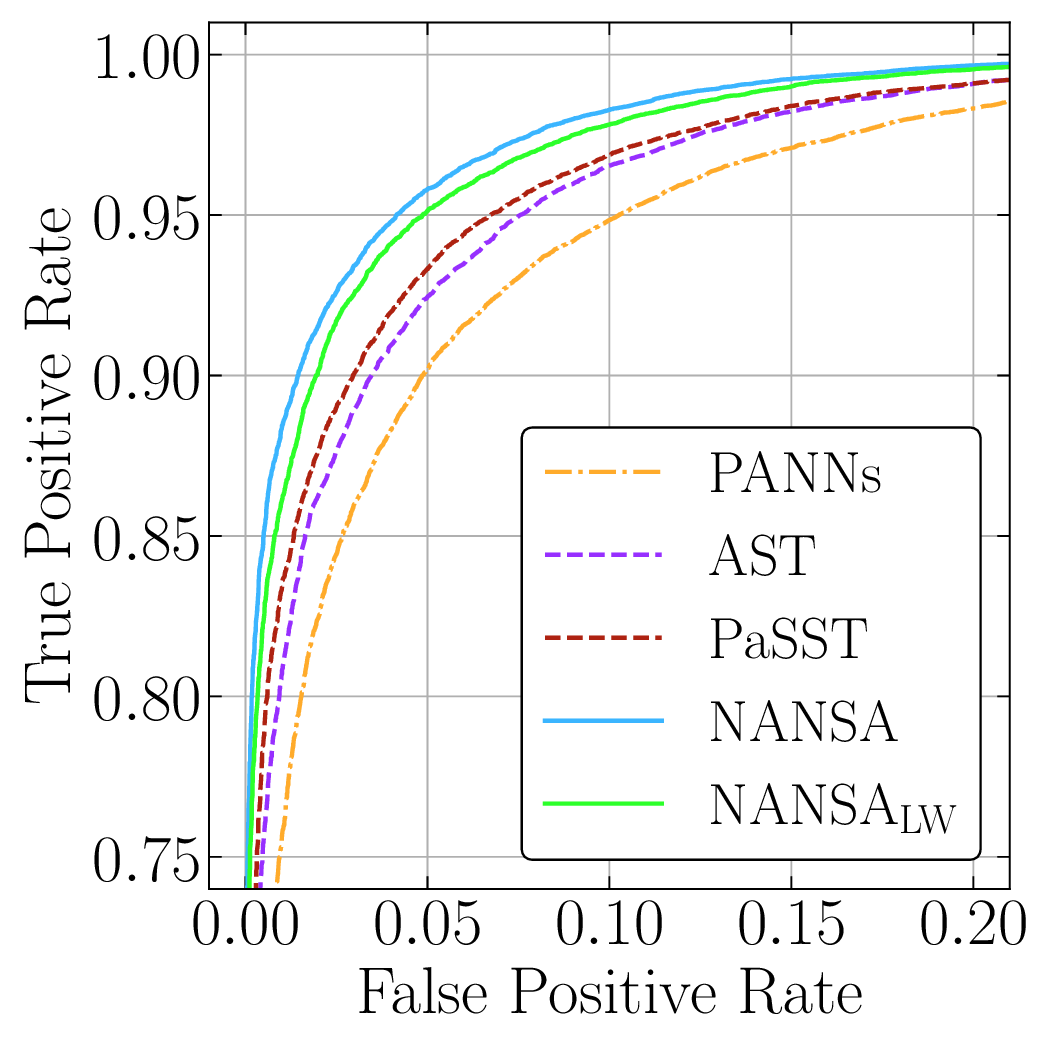}
    {\renewcommand{\tabcolsep}{0.3mm}
    \renewcommand{\arraystretch}{1.8}
    \fontsize{6.9pt}{6.9pt}\selectfont
    \begin{tabular}{l c c c }
    \vspace{-5.2cm} \\
    \hhline{~---}
     & \multicolumn{3}{|c|}{ \bf AUC }  \\ \hhline{~---} 
     & AudioSet & DCASE & FSD50K  \\ \hhline{----} 
    PANN  & 0.973 & 0.980 & 0.982  \\ \hhline{~~~~} 
    AST   & 0.980 & 0.986 & 0.987  \\  \hhline{~~~~} 
    PaSST & 0.979 & 0.986 & 0.989  \\  \hline
    NANSA & \bf 0.989 & \bf 0.996 & \bf 0.993 \\  \hhline{~~~~} 
    NANSA\textsubscript{LW} & 0.986 & 0.995 &  0.992 \\ \hhline{~~~~}
    \hline
    \end{tabular}
    }
    \vspace{-0.5cm}
    \caption{ROC curves and Area Under Curve (AUC) for acoustic non-stationarity assessment.}
    \label{fig:rocs_complete}
    \vspace{-0.5cm}
\end{figure}

Results for the proposed NANSA and NANSA\textsubscript{LW} models are also shown in Table~\ref{tab:all_scores}.
These models are specifically designed for non-stationarity assessment and consistently outperform the baselines across all metrics, while being far more efficient in terms of model size and computation.
On AudioSet, NANSA achieves the highest accuracy—1.8 percentage points higher than AST.
For DCASE and FSD50K, it yields substantial EER reductions of 49.1\% and 20.8\%, respectively, relative to the best baseline.
NANSA also achieves a 30.9\% higher F1-score than PaSST on the DCASE dataset.

Similar trends are observed for the lightweight NANSA\textsubscript{LW}.
On average, both NANSA variants achieve over 95\% accuracy and EER values below 5\%, demonstrating strong reliability in acoustic non-stationarity classification.
While NANSA\textsubscript{LW} slightly underperforms compared to the full NANSA model, it consistently surpasses all baselines across metrics and datasets.
For instance, on AudioSet, NANSA\textsubscript{LW} achieves an EER of 6.73, representing an 15\% reduction compared to the AST baseline.

For the ablation study, the impact of the ANS Encoder module is also summarized in Table~\ref{tab:all_scores}. 
Preliminary experiments explored $\beta_{FC} \in \{1.5, 2, 4, 6\}$, with $\beta_{FC}=4$ selected due to slightly better performance. Removing the ANS Encoder increases average EER by 10.5\% and 12.5\% for NANSA and NANSA\textsubscript{LW}, respectively, with the latter being more affected due to its smaller model capacity.

\begin{table}[t!]
\centering 
\renewcommand{\tabcolsep}{1.0mm}
\renewcommand{\arraystretch}{1.2}
\vspace{-0.2cm}
\caption{Processing time comparison between INS original algorithm and data-driven HLC-based models. Gray values (xN) indicate the improvement factor over INS.} 
\fontsize{7.7pt}{7.7pt}\selectfont
\begin{tabular}{|c| c c c c c|} \hline 
    \multicolumn{6}{|c|}{ Processing Time (ms)}  \\ \hline
  INS & PAANs & AST & PaSST & NANSA & NANSA\textsubscript{LW}  \\ \hhline{------} 
12597.1$\pm$25.3  & 32.0$\pm$4.2 & 133.0$\pm$8.2 & 115.5$\pm9.7$ & \bf 27.3$\pm$1.3 & \bf 3.2$\pm$0.1  \\  \hhline{~~~~~~} 
\color{gray} (x1)  & \color{gray} (x394) & \color{gray} (x95) & \color{gray} (x110) & \bf\color{gray} (x466) & \bf \color{gray} (x3957)  \\  \hline
\end{tabular}
\label{tab:time_comparison}
\vspace{-0.5cm}
\end{table}

Fig.~\ref{fig:rocs_complete} shows the ROC curves and corresponding AUCs for each dataset.
In all scenarios, NANSA and NANSA\textsubscript{LW} curves are closest to the ideal operating point $(0, 1)$.
Accordingly, the proposed models achieve the highest AUCs, further confirming their efficacy in non-stationarity classification.

It is remarkable that all acoustic models attained consistent classification results via HLC algorithm and therefore can be adopted as a solution to overcome INS resource intensive issues.
In Table~\ref{tab:time_comparison}, it is shown the comparison between the inference time of HLC-trained models with the original INS statistical framework.
All models significantly reduce processing time compared to INS, thanks to HLC-based training.
Notably, NANSA and NANSA\textsubscript{LW} are approximately 500 and 4000 times faster than the traditional INS approach.
The time efficiency gain confirms the effectiveness of non-stationarity assessment for HLC-trained models in both large-scale and resource-constrained devices.
One drawback of HLC is the loss of scale-relative INS information. 
As future research, we intent to overcome this with multi-task learning strategies.

\vspace{-0.2cm}
\section{Conclusion}
\vspace{-0.1cm}
This work addresses the challenge of objective and computationally feasible acoustic non-stationarity assessment.
The HLC was introduced as a novel labeling algorithm that enables supervised learning models to replace traditional INS-based evaluations.
We validated HLC across multiple datasets and architectures, and proposed a dedicated model named NANSA, which consistently outperforms state-of-the-art baselines.
Extensive experiments demonstrated that HLC-trained models provide reliable, scalable, and fast solutions for non-stationarity estimation, overcoming the computational limitations of conventional INS framework.

\bibliographystyle{IEEEbib}
\bibliography{mybib.bib}

\end{document}